\documentclass{article}

\usepackage{arxiv}

\usepackage[utf8]{inputenc} 
\usepackage[T1]{fontenc}    
\usepackage{hyperref}       
\usepackage{url}            
\usepackage{booktabs}       
\usepackage{amsfonts}       
\usepackage{nicefrac}       
\usepackage{microtype}      
\usepackage{graphicx}
\usepackage{float}
\usepackage{caption}
\graphicspath{ {./} }

\title{The tradeoff between the utility and risk of location data and implications for public good}

 \author{
   Dan Calacci \\
   MIT Media Lab\\
   \texttt{dcalacci@media.mit.edu} \\
   \And
  Alex Berke \\
   MIT Media Lab\\
   \texttt{aberke@mit.edu} \\
   \And
  Kent Larson \\
   MIT Media Lab\\
   \texttt{kll@media.mit.edu} \\
   \And
  Alex 'Sandy' Pentland \\
  MIT Media Lab\\
  \texttt{sandy@media.mit.edu} \\
 }

\begin{document}
\maketitle
 
\begin{abstract}

High-resolution individual geolocation data passively collected from mobile phones is increasingly sold in private markets and shared with researchers. This data poses significant security, privacy, and ethical risks: it's been shown that users can be re-identified in such datasets, and its collection rarely involves their full consent or knowledge. This data is valuable to private firms (e.g. targeted marketing) but also presents clear value as a public good. Recent public interest research has demonstrated that high-resolution location data can more accurately measure segregation in cities and provide inexpensive transit modeling. But as data is aggregated to mitigate its re-identifiability risk, its value as a good diminishes. How do we rectify the clear security and safety risks of this data, its high market value, and its potential as a resource for public good? We extend the recently proposed concept of a tradeoff curve that illustrates the relationship between dataset utility and privacy. We then hypothesize how this tradeoff differs between private market use and its potential use for public good. We further provide real-world examples of how high resolution location data, aggregated to varying degrees of privacy protection, can be used in the public sphere and how it is currently used by private firms.
\end{abstract}


\section{Introduction}

There is a flourishing data economy that many of us, simply by owning and using smartphones, are participating in.  Data collected from smartphones detailing the times and locations of where individuals have been is collected at a scale and a level of precision\footnote{Location Based Services data can be precise up to 10 meters.} that is unprecedented.  This data jeopardizes individual privacy and security, and the way it is collected often circumvents users’ awareness or explicit consent.  This high-resolution location data presents value to private firms who are interested in when and where people have been.  However, it also presents potential value for the public from whom it was obtained, and this value has not been fully realized. In this paper, we evaluate the balance between the value of this data to the private market and the public interest, and its implications for individual privacy and personal risk.

In doing so, we challenge existing measures of risk and utility that are commonly used in the computational literature.  We argue that these measures tend to remove the social context from location data and its use in order to make statistical computation feasible.  Yet it is only with the context of how these datasets are used that the value and risk they present to organizations and people can be fully understood.

\subsection{The large scale collection of modern mobility data}
A variety of smartphone applications collect our location data for reasons that are both benign and surreptitious.  Often, this data serves the application’s function.  For example, weather apps use our location to provide local weather forecasts\footnote{The weather app WeatherBug fell under scrutiny when it was discovered that it was owned by a data broker, GroundTruth, which was also collecting and using the location data to help its clients better serve personalized ads \cite{valentino-devries_singer_keller_krolik_2018}.}, while mapping applications compute distances and directions relative to where we are.  News apps might use location data to serve nearby news, while sports apps report on local teams\footnote{Sports app theScore was also found to pass the precise location coordinates of its users to advertising and location companies \cite{valentino-devries_singer_keller_krolik_2018}.}.
	
These applications also commonly share this location data with third parties who are in the business of collecting user data, analyzing it for market insights, and selling it to help businesses with targeted advertising campaigns and a range of other purposes.  These data brokers often pay application developers to include code in application software (through e.g. a “Software Development Kit”, or SDK”) \cite{times_2018_1} to collect location data, in the background, without the user’s full knowledge.  The companies that develop these SDKs and collect location data at scale are referred to as “Location Based Services” (LBS) companies, and the aggregate data they collect is referred to as “LBS” data. A New York Times investigation showed that many applications that use these SDKs do not properly notify users of the full extent to which their location data will be used or shared. They also may bury that information within privacy policies that are difficult to find \cite{valentino-devries_singer_keller_krolik_2018}.

The different applications on our phones may only collect our location data in limited intervals, such as when they are in use or running in the background, depending on the privacy settings we have set.  However, the aggregate of this data reported from these various applications is what the data brokers obtain. The resulting datasets detail the daily movements of hundreds of millions of people. They pose significant privacy and security risks for the individuals from whom the data was collected, yet they are available for purchase and shared with researchers and other groups. We use the example of LBS data as a case study, highlighting the lack of sufficient anonymization in these sensitive user location datasets that are available for sale by private firms. LBS data also provides insights into the location data market and how this data can be used, for better or worse.

\subsection{Nature of the mobility data}
The data is a set of timestamped latitude, longitude coordinates representing where and when each of the people from whom it was collected have been. This type of data is referred to as “trajectory data” because it allows an analyst to follow or probabalistically infer an individual’s path, or trajectory, from point A to point B.  In this data we can observe where a person been, how long they stayed there, and the route they took to get there.  The velocity of an individual at any point in a trip can also be extracted from trajectory data, and used to classify which mode of transport they were likely taking \cite{zheng_2015}.  
This data can be used to observe and model a population’s mobility patterns when analyzed at an aggregated level.  At a more granular level, it can be used to observe an individual’s daily patterns, and even predict or help influence where they might go next. For example, the home locations of the majority of users represented in LBS datasets can be inferred by observing where they regularly stay between the hours of 8pm and 4am.

\subsection{Mobility data and privacy implications}
The data, as it is provided, is pseudo-anonymized so that random user identifiers are attached to users’ points instead of explicit names or identities.  However, this crude de-identification procedure does not grant individuals privacy and their identities can easily be recovered \cite{narayanan2014no}.  Previous work has shown that only a few timestamped geolocation points are needed to uniquely identify users \cite{de2013unique}.

While the  large-scale collection of location data via smartphone applications may be relatively new, the large-scale collection of spatial trajectory data from mobile phones via telecoms is not.   
Telecom companies around the world have long been collecting and storing historical call detail records (CDRs) that contain location and communication data about their customers.  CDRs contain metadata from mobile phone use, including which antenna a mobile phone communicated with and when.  Because phones communicate with the nearest antenna, an antenna's location is a proxy for the general location a phone's user was in.

This personal location data from CDRs has been used by prosecutors, such as those at the Manhattan District Attorney’s office, to suggest defendants were in the same locations as crimes that took place, and help argue their cases \cite{10107249385974248_2017}.   Prosecutors may have privileged access to de-anonymized CDRs, but sensitive personal location data can still be recovered from the pseudo-anonymized CDRs that are more generally available to researchers and companies.

A seminal 2013 study of CDRs from 1.5 million users whose data had been scrubbed of their identities (“de-identified”) showed the ease with which users could be re-identified, due to the highly unique nature of individual “mobility traces” \cite{de2013unique}.  Just four randomly selected spatio-temporal points were enough to uniquely identify 95\% of users in the dataset.  This means that with knowledge of just four places a person has been, and when they were there, a person could be identified in this dataset with a 95\% likelihood because no other person was at all four of those places at those times.  This likelihood increases with the number of known spatio-temporal points.  Once re-identified, all other points in the dataset associated with a user become available and known. 
This significant privacy risk is even more pronounced for the modern datasets collected from smartphone applications and stored by LBS firms.  Smartphone applications collect longitude and latitude coordinates which are much more precise than the data in CDRs, which is based on cell phone antenna locations.
The few necessary points needed to re-identify someone are easy to come by.  Home addresses can be looked up in public records for homeowners.  Where someone works, or a coffee shop or event they have gone to, can be discovered anecdotally.  Such data points are even more commonly accessible through social media or other public datasets.  With a few geo-tagged tweets, one could re-identify a Twitter user in a location dataset.
	
Even without being identified, location data poses a security risk for individuals because of its ability to expose sensitive places they have been. The Electronic Frontier Foundation published a list of such sensitive places and the risks associated with their knowledge \cite{eff_2011}.  Examples include abortion clinics, political rallies, gay bars, mosques, and other places of worship. The fact that a person has been to one of these places may be more relevant than that person’s identity. Their pseudo-anonymized user ID can then be followed through a location dataset to the other places they go, their home can be determined, and they can be targeted.

\subsection{Value of mobility data}
The ability to follow individuals from location to location, whether or not their name or identity is  known, is also what makes this data so valuable.
As early as the mid-2000s, researchers and private firms expressed interest in extracting value from the ubiquitous data source presented by CDRs. By 2004, patents detailing location detection of normal cellular phones began to appear \cite{koorapaty2004efficient}. These methods were presented as ways for telecom companies to charge users based on location, or create value-added services on top of existing telecom services \cite{awada2002coupon,mcdowell2006proximity}. Shortly after, researchers also began to investigate the use of CDRs for predictive modeling and population-level measurement \cite{ratti2006mobile,calabrese2010geography}. 

More modern and precise location data from smartphones presents even more value realized through targeted ad campaigns, both in theory and in practice. Market research has shown that businesses can reap higher returns on their ads if they are targeted to reach people when in crowded places \cite{andrews2015mobile}.  In practice, location data has been used to target ads for personal injury lawyers to people in emergency rooms, and used to target ads from a Christian pregnancy counseling and adoption agency to people who entered Planned Parenthood clinics\footnote{For which the responsible advertising agency was sued by the Massachusetts Attorney General \cite{MA_AG_CopleyAdvertising}.} \cite{MA_AG_CopleyAdvertising} \cite{allyn_2018}.  

More advanced targeted marketing campaigns often result from closer partnerships between data brokers and their client businesses.  For example, the data broker GroundTruth used location data to advance their client Timberland’s ad campaign by identifying audiences likely to buy Timberland footwear.  They did so based on people’s proximity and recent visits to Timberland stores, competitor footwear stores, and whether they had recently visited an outdoor activity location \cite{groundtruth_timberland}.

While the value of location data has been widely recognized for private firms, it also presents significant value for the public.  GroundTruth, the same data broker that used location data to power Timberland’s ad campaign, also uses location data to send AMBER alerts to users in areas determined by US law enforcement agencies \cite{groundtruth_amber}.  These notifications alert recipients about missing children in order to more efficiently crowdsource information needed for their rescue.

Moreover, the collection, analysis, and use of the public’s location data is not a new practice for  government organizations.  Governments have even facilitated it.  In the case of the United States, the emergency services initiative Enhanced 911, created by the FCC, required telecom companies to be able to locate users within a 150 meter radius 97\% of the time \cite{spinney2003mobile}.

Governments also dedicate large amounts of resources to conduct surveys, such as the U.S. Census and the National Household Travel Survey. Surveys provide important information about the population to government organizations and researchers, to help better plan for the future and allocate resources.  For example, the National Household Travel Survey produces a mobility dataset that has been used for the purposes of traffic safety, congestion, the environment, energy consumption, demographic trends, bicycle and pedestrian studies, and transit planning \cite{nhts_2017_compendium}.

Collecting this data is expensive and difficult, so it is collected infrequently.  Survey responses can then only provide snapshots in time of the population and the collected data only reflects the answers respondents were willing to give, and are subject to recall bias \cite{oliver2015mobile}.  In comparison, data collected from mobile phones reflects what the “surveyed” populations actually did rather than what they stated they did, and the data is real-time rather than an outdated snapshot.  Researchers have shown that this inexpensive and readily available data source can be used to supplement surveys and fill in gaps between surveys that are only conducted every several years \cite{deville2014dynamic}. 

Within the U.S., government, agencies have become increasingly aware of how location data from mobile phones can improve state services.  In 2019, the state of Maine purchased a subscription to a data stream from the company Streetlight Data, a location data aggregation company, to improve the state’s transportation planning and infrastructure \cite{maine_2019}.
Mobile phone data can also provide important population data for countries where surveys are not conducted at all, or provide information in moments of crisis \cite{deville2014dynamic, jahani2017improving}.  GPS data collected at scale has been used to help measure and understand the spread of disease and can be used to develop more effective natural disaster strategies \cite{nakajima2007disaster, vazquez2013using}.

\section{The Risk-Utility Tradeoff}
There is a clear tradeoff between the utility of location data to both private firms and the public good, and the privacy risk it presents to individuals.  How should this data be structured to preserve user privacy while still yielding useful information? There are a variety of strategies and technical solutions to address this problem, and a sizable body of literature quantifying their impact on utility and individual privacy risk.  Most of these works address strategies where data providers aggregate data to larger spatial or temporal resolutions, which can make it harder to re-identify individuals while still providing value.

Consider a LBS dataset in its standard form. The same user ID is associated with each timestamped location point for a given user (Table \ref{tab:agg1}), providing for a more precise analysis of user trajectories. Coarsening this data spatially might mean replacing a user’s precise GPS coordinates with the city block the user was on, or a statistical area such as a census block group. Providers can also coarsen data along the dimension of time, where time might be stored in 1-hour or 6-hour increments, rather than at the minute (or millisecond) level of accuracy available to LBS firms. Simple coarsening of spatial or temporal resolution can increase the number of points needed to re-identify someone in a large dataset \cite{de2013unique,zang2011anonymization}, and therefore reduce risk. But this reduction is limited.

The same seminal 2013 study which showed that just four spatio-temporal points were enough to uniquely identify 95\% of the 1.5 million users in a CDR dataset also analyzed the relationship between re-identifiability and data resolution.  The authors found that “the uniqueness of mobility traces decays approximately as the \( \frac{1}{10} \) power of their resolution”\cite{de2013unique}. Other studies find that reliable anonymization is only realized when data is coarsened so much as to be of little value, such as aggregating spatial resolution to the level of cities \cite{zang2011anonymization}. These key insights show that while reducing data resolution does increase privacy, the privacy gain is slow, and “even coarse datasets provide little anonymity\cite{de2013unique}. Still, these “legacy solutions”, often variations of the seminal k-anonymity algorithm introduced in 1998, are the de-facto current standard for mitigating risk in location datasets \cite{gramaglia2015hiding,de2018privacy}.

The problem of balancing personal risk with utility in behavioral datasets, and location datasets in particular, isn’t new. There are sizable computational literatures that detail both re-identification and anonymization strategies for CDRs and location data, and they often directly or indirectly imply a relationship between utility and risk \cite{barak2016anonymizing, tu2017beyond, hardjono2019trusted, gramaglia2014anonymizability, de2013unique,zang2011anonymization}.  To formalize the concepts of risk and utility, researchers persistently define both in ways that strip them of their social contexts. Risk and utility are often defined as measures of the data itself, or as the abstract predictive power of state-of-the-art models trained on it. In the case of risk, these measures usually amount to the average quantity of data from a secondary source someone would need to identify an individual in a dataset. These definitions avoid the social context in which data is collected or shared, and so may lead to a false sense of security about its release. As an example, take the relative risk of membership in a mobility dataset for a U.S. citizen and a non-citizen immigrant. Their contextual risk is radically different. The U.S. citizen may never be identified or targeted, while the immigrant has a very real risk of being targeted and deported by the state using location information. U.S. Immigration and Customs Enforcement (ICE) already uses license-plate reader data to target immigrants for deportation, and has discussed using mobile location datasets to do the same \cite{aclu2019}. Conversely, in the context of utility, one must ask: “utility for whom?”. What utility does the immigrant community gain through the release of such data? 

Another limitation of considering risk and utility computationally is that it risks mischaracterizing both as static attributes of a specific dataset, rather than a dynamic quantity. Both the risk and utility of a dataset should be considered relative to the state of adversarial technologies, the ease of acquiring external data about an individual in the set, as well as the contextual social and political use cases of the data itself. As technology has advanced, new techniques for re-identifying individuals in aggregated or ‘privacy-preserving’ data have been developed. In 2017, a group of researchers published a technique to re-create individual trajectories from aggregated location data that had been considered highly protective of individual privacy by the research community \cite{xu2017trajectory}. We expect such techniques to only develop and become more widespread in the future. At the same time, external data that can aid in re-identifying individuals has become increasingly accessible. For example, social media posts on public networks such as Instagram and Twitter are commonly geotagged with an individual’s location.  Even without geotagging, an individual’s location can be inferred through images or local information in their post.

\begin{figure}
  \centering
  \includegraphics[scale=.5]{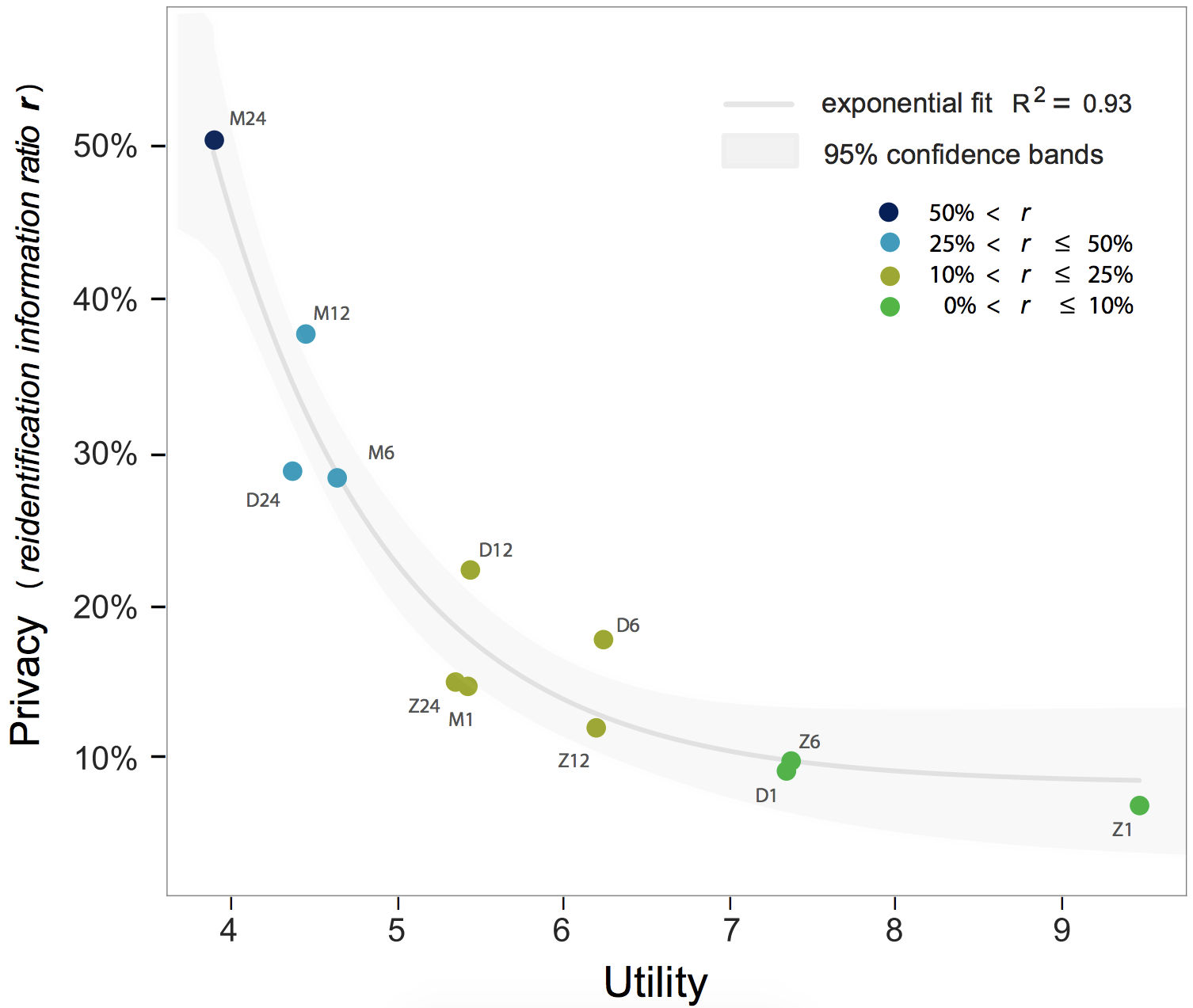}
  \caption{Curve developed by Noriega-Campero et. al: Privacy-Utility trade-off in mobile phone data. It plots utility vs. re-identification risk in mobile phone data, across spatial and temporal granularities \mbox{\{ZIP, District, Municipality\} x \{1h, 6h, 12h, 24h\}}. The more useful the dataset, the less auxiliary information is needed to re-identify its individuals. Conversely, while data generalization increasingly hinders re-identification, it strongly diminishes datasets’ value \cite{noriega2018mapping}.  Note that we switch the axes to plot utility on the vertical axis in our extension of this curve.}
  \label{fig:curve1}
\end{figure}

Given these limitations, we aim to provide workable, contextual examples of how the trade-off between ‘risk’ and ‘utility’ has appeared in the real world without formal or computational definitions, and speculate on how this trade-off might present itself in the future. Instead of formally quantifying risk, we describe varying aggregation strategies applied to the same hypothetical LBS dataset that might incrementally mitigate its level of risk.
For each aggregation case, we describe brief specific examples of current or potential uses in two different contexts. First, we detail an example of how private firms currently use or might use the data at that aggregation level. Second, we describe how such data is currently used for ‘public interest’ projects and can serve the public good. In both cases, we provide brief analyses of the risk and utility these examples afford in an attempt to bridge the gap between computational measures of privacy and risk of the data itself and the social contexts of its possible use.

\begin{figure}[H]
  \centering
  \includegraphics[scale=.75]{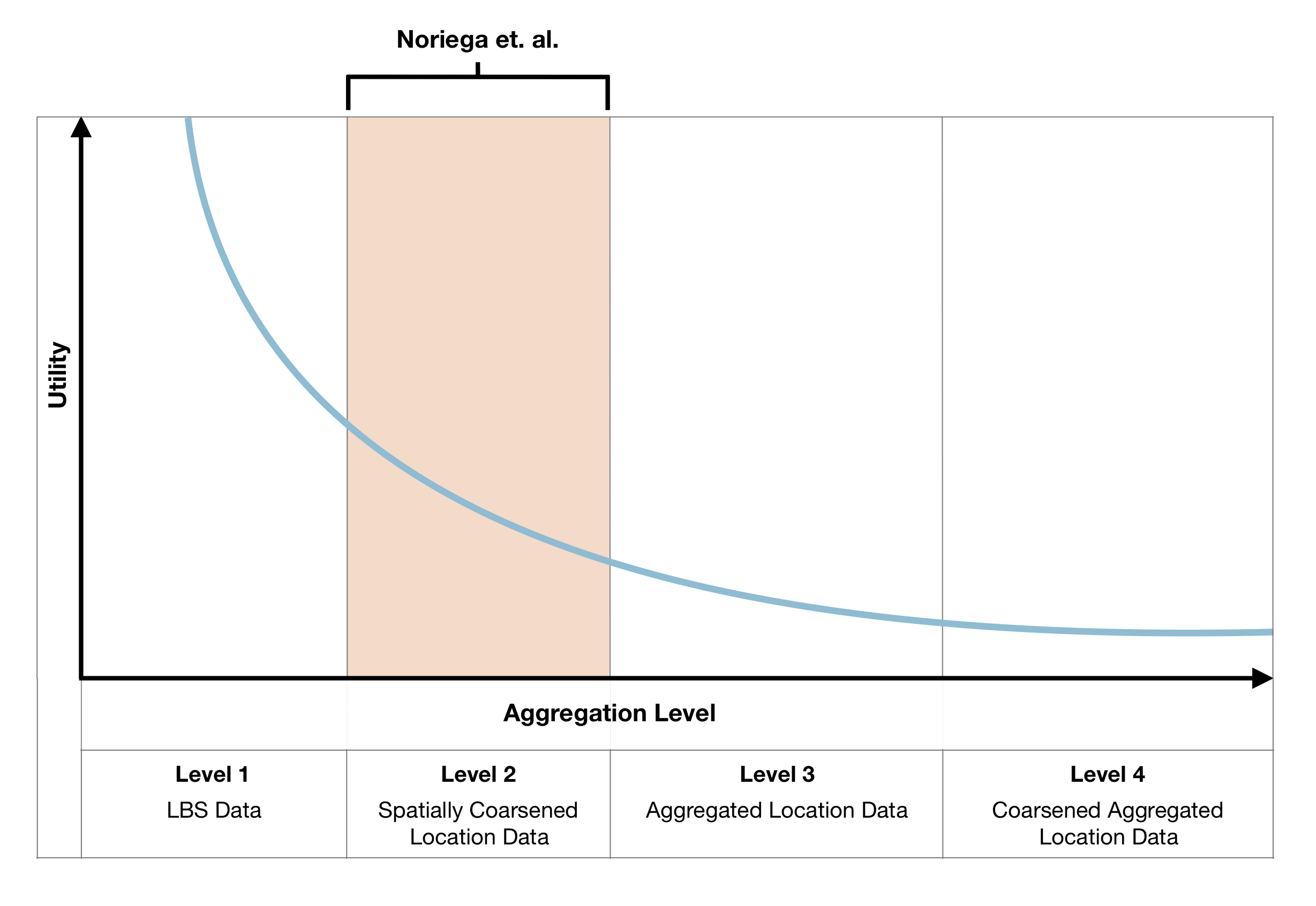}
  \caption{Our speculative extension of the utility-privacy curve presented by Noriega-Campero et. al.  We plot utility on the y-axis and shade the portion of the curve discussed by Noriega-Campero.  We extend the curve by considering additional levels of data aggregation, and the associated uses and risks for data at each aggregation level.  Each horizontal segment, which we demarcate by enumerated ’levels’, corresponds to a different aggregation strategy we discuss.  Like Noriega et. al., our curve shows that the utility of data decreases as its level of aggregation increases.}
  \label{fig:curve2}
\end{figure}

To help contextualize the relationship between our analysis and the computational literature on risk and privacy, we borrow the idea of a utility-privacy curve (figure \ref{fig:curve1}), recently explored by Noriega-Campero et. al \cite{noriega2018mapping}. For different aggregation levels of CDR data, the authors compute the re-identifiability risk of the data, and quantify its ‘utility’ by surveying experts in CDR data use. Our aggregation cases differ from the ones presented by Noriega-Campero in two key ways. First, we consider high-granularity GPS data like that shared by LBS firms rather than CDR data. The most granular data we discuss carries higher risk and different utility than the most granular data discussed by Noriega-Campero. Second, we consider alternative forms of aggregation that are considered less risky and less granular than those considered on the curve by Noreiga-Campero et al. Figure \ref{fig:curve2} illustrates how a speculative curve that includes our examples might extend the curve as presented by Noriega-Campero. We extend the curve by adding examples both in regions of lower and higher aggregation.

\subsection{Location Based Services Data}

\subsubsection{Level of Aggregation}
Our first case considers the kind of location data shared by LBS companies in the data market today. This data can contain timestamped latitude and longitude points for each user in the dataset, accurate to within 10 meters and a few seconds. It is highly re-identifiable, and often allows firms to infer the precise positions of a user’s home, work, and leisure locations. Table \ref{tab:agg1}  shows an example set of location data from one user.

\begin{table}[h]
  \centering
    \caption{Example rows for LBS Location Data. Details specific user ID, the latitude, longitude, and time (at the resolution of minutes) of each data point.}
  \begin{tabular}{llll}
    User ID     & Latitude     & Longitude & Time \\
    \midrule
    a7e9a3655c3f9da6532465a721a91f9e & 41.83  & -72.237 & 1:00pm \\
    a7e9a3655c3f9da6532465a721a91f9e & 41.84  & -72.243 & 1:01pm \\
    a7e9a3655c3f9da6532465a721a91f9e & 41.83  & -72.251 & 1:01pm \\
    a7e9a3655c3f9da6532465a721a91f9e & 41.82  & -72.257 & 1:03pm \\ \bottomrule
  \end{tabular}
  \label{tab:agg1}
\end{table}

\subsubsection{Market Utility}
The market utility of data like this is hard to overstate. Individual, highly precise geolocation data is part of a new targeted geo-marketing industry, and the “holy grail” of advertising \cite{laurenjohnson2017}. One use of this data is to link contextual behaviors to advertising targets.  For example, businesses can target ads to individuals based on their previous visits to a store, a competitor’s store, or their patterns of passing a storefront.  Timberland did just this by working with LBS data provided by GrouthTruth in an attempt to sell more footwear \cite{groundtruth_timberland}.
Combined with geofencing strategies, user location data can be used for even more proactive marketing. With geofencing, target audiences can be identified based on their real-time location and delivered an ad at the opportune moment, such as the case of targeting ads for personal injury lawyers to people in emergency rooms \cite{allyn_2018}.
Target audiences can also be identified based on their attendance to an event.  In an interview, a Chief Marketing Officer explained that their firm “...used geofencing for an entertainment client to target people at Comic Con because we knew our show’s audience was there. That data drove activation, audience understanding and future targeting”. By tracking people between events and places, including those seemingly unrelated to a product or service, marketing firms are able to move beyond market segmentation and deliver ads to individuals, based on their real-time behavior\cite{laurenjohnson2017}.

\subsubsection{Public Utility}
At the same time, highly individual GPS trajectory data can have a large amount of public utility. In the United States, when people in emergency situations call 911, their GPS coordinates are automatically provided to emergency dispatchers.  This was made possible by the enhanced 911 initiative (E-911) in order to send callers help in a more timely manner.
GPS data has also been utilized for public health.  A study in Peru used GPS to track individuals who had been exposed to the Dengue virus in order to better understand its risk of transmission.  Individualized location tracking of this kind has shown its utility in helping researchers and government agencies monitor the spread of infectious disease in efforts to protect the public from widespread epidemics \cite{vazquez2009usefulness}.

\subsubsection{Risk}
Given that the coarse CDR data can be used to re-identify users with just 4 points and reconstruct their trajectories, data that records latitude and longitude coordinates is even riskier.  From this data, precise locations visited and home addresses can be inferred for re-identified users.
This risk is not isolated to usage by “bad actors” who access the data through illegal means. In many cases, the re-identifiability of the data is it’s selling point: the “bad actors” that cause personal risk are the legal users of the data. Police in the U.S. have used GPS data from mobile phones to identify and arrest suspects without a warrant, a practice that is now illegal\cite{valentino2019devries}. Bounty hunters and skip-tracing firms seemingly use this data to locate people with no apparent oversight for low fees \cite{cox2019}. The risk that these uses bring to certain populations can be devastating. Stalkers and domestic abusers are known to use malware apps to track their targets and spouses, and bounty hunters have admitted to using their data subscriptions to track their own partners\cite{cox2019,franceschi2017}. The clearest use cases for this data outside of city and emergency planning and marketing are products that allow citizens and the state to target and trace individuals more effectively than ever before, indicating a high risk of abuse.

\subsection{Spatially coarsened location data}
\subsubsection{Level of Aggregation}
This case of coarsened LBS data still refers to individual trajectory data, but with a lower spatial resolution than its original, high-resolution form. Records are still identified by a unique user ID, but instead of a spatial resolution at the order of meters, locations are identified by their Census Block Group.  Table \ref{tab:agg2} shows an example of this data with a temporal resolution of one minute.

\begin{table}
 \caption{Example of spatially coarsened location data. Each row details the census block group a specific user was in at a particular time, at the minute resolution.}
  \centering
  \begin{tabular}{llll}
    User ID     & Census Block Group & Time \\
    \midrule
    a7e9a3655c3f9da6532465a721a91f9e & 250173164003 & 1:00pm \\
    a7e9a3655c3f9da6532465a721a91f9e & 250173165003 & 1:01pm \\
    a7e9a3655c3f9da6532465a721a91f9e & 250173165003 & 1:15pm \\
    a7e9a3655c3f9da6532465a721a91f9e & 250173164003 & 1:32pm \\ \bottomrule
  \end{tabular}
  \label{tab:agg2}
\end{table}
 
This level of granularity can be considered similar to that of CDRs, whose use cases have been well studied and documented.  Mere observation of this data reveals the movements of individuals between neighborhoods.  With simple tooling, mobility data at this resolution is detailed enough to describe traffic flows, geographic areas that an individual frequents, and a person’s daily routines \cite{blondel2012data}.   CDRs with this type of location data have been used in machine learning models to correctly infer the professions and unemployment status of individuals, as well as other socio economic characteristics \cite{sundsoy2016estimating}.

\subsubsection{Market Utility}
While businesses can no longer identify specific store visits using this coarser data, they can still perform targeted marketing.  For example, “outdoor enthusiasts” can be identified by tracking users who spend more time in state parks or rural areas.  L.L. Bean used location data in 2019 to target “outdoor enthusiasts” for online ads about their summer product line \cite{llbean2018}. 

Identifying and targeting “outdoor enthusiasts” is an example of using market segmentation to drive profits.  Market segments are also commonly divided based on characteristics such as gender, age, or profession. These characteristics can be inferred from the coarse mobility data provided in CDRs, showing how this data can further contribute to the consumer revenue streams of private firms.

Data from CDRs has also been used to predict an individual’s likelihood to make loan payments, providing banks and creditors with more information on whether to extend a loan.  This can be particularly useful to financial institutions in developing countries where many households lack formal financial histories \cite{bjorkegren2018behavior}.

\subsubsection{Public Utility}
LBS data aggregated at this level represents individual trajectory data similar to what is already collected by government surveys.  The United States’ National Household Travel Survey samples the population for data on their routine trips; the data serves a variety of public analyses and projects.  For example, travel behavior is used for traffic prediction to help city transportation engineers better plan roadways and mitigate congestion \cite{cityscience_andorra_traffic}.  Aggregated LBS data can supplement travel surveys with data collected at a scale much greater than survey samples.  And this data can be collected in real-time instead of the limited snapshots provided by surveys, providing a more dynamic source of information for city governments.

This aggregregated LBS data can be used to supplement other surveys as well.  The Current Population Survey (CPS) is a monthly survey of about 60,000 U.S. households conducted on behalf of the Bureau of Labor Statistics in order to estimate the unemployment rate and other market indicators \cite{us_census_bureau_2018}.  Since researchers have shown that CDRs can be used to infer people’s employment status \cite{sundsoy2016estimating}, this data could be used to supplement such surveys and monitor changes in employment status, or provide for countries where labor survey data is not easily collected.

Individual LBS data at this level of spatial aggregation can also be used to serve individuals timely alerts on behalf of the public.  In the United States, the Federation of Internet Alerts subscribes to a LBS company in order to serve AMBER Alerts and severe weather warnings to people based on their location.   According to the U.S. Office of Justice Programs, 957 children have been rescued due to the AMBER alert system \cite{amber_alert_stats}.

\subsubsection{Risk}
The re-identifiability risk in this data is high.  Researchers have shown that only four spatio-temporal points are needed to re-identify 95\% of users in such a dataset.  This data no longer reports people’s precise locations, making it more difficult to infer home addresses or the sensitive places they may have visited.  However, understanding the daily mobility traces of individuals still provides valuable information to skip-tracing firms and law enforcement agencies.

This data has been particularly useful to law enforcement agencies, who do not need to do the extra work to re-identify individuals because they can ask for a person’s data directly from providers.

Using CDRs to track suspects’ movements and using their locations to implicate them at the time of trial has become common. Outside software firms have begun developing and marketing software to analyze CDRs on behalf of police and other investigators (e.g. CellHawk by Hawk Analytics\footnote{http://www.hawkanalytics.com}).   The software maps out user locations from their uploaded CDRs and “also reveals general GPS attuned information about the target phone, such as the most frequent locations visited, where the target spends most of their time, where they work, and where they sleep” according to a marketing blogpost for one such software provider \cite{policeone2017}.  The risk inherent to this data was highlighted in 2018 when the US Supreme court ruled in \textit{Carpenter v. United States} that police must first obtain a warrant before accessing a suspect’s CDRs \cite{carpenter.v.united.states_2018}.  In their decision, the court wrote that cell phone location records “give the Government near perfect surveillance and allow it to travel back in time to retrace a person’s whereabouts, subject only to the five-year retention policies of most wireless carriers.”  

While requiring a warrant for police to analyze CDRs may seem like a prudent measure, the ease of obtaining a warrant varies depending on the local judiciary.  In addition, the \textit{Carpenter v. United States} opinion was narrow and did not extend to matters of national security, such as investigations by the U.S. Immigration and Customs Enforcement (ICE).  ICE, which commonly pursues undocumented immigrants for deportation, may continue with limited oversight to use cell phone location data to track and arrest potentially vulnerable individuals who have little opportunity to contest their cases.

Trajectory data also presents risk when used for predatory targeted marketing.  The ability to infer socioeconomic characteristics for market segmentation or employment status to estimate market indicators from an individual’s CDRs may be of utility when used in good faith.  However, this information can also be exploited to target ads to vulnerable individuals.

For example, for-profit colleges, known for misleading students \cite{lewin_2010}, have targeted and recruited single mothers \cite{institute_for_womens_policy_research}.  Payday loan agencies, known for their business model of sending customers into spirals of debt \cite{paydayloans_newhire_training}, often target low-income or recently unemployed people \cite{opencashadvance}.

\subsection{Aggregated Location Data}
\subsubsection{Level of Aggregation}
This level of aggregation no longer represents individual trajectory data. Each user is now identified by an identifier corresponding to their home census block group or other statistical area, rather than by a unique ID (Table \ref{tab:agg3}). The recorded locations might be highly precise, but they are not directly tied to the users that generated them.

\begin{table}
 \caption{Example aggregated location data. Each row details the latitude, longitude, and time for a user from a given census block group.}
  \centering
  \begin{tabular}{llll}
    Census Block Group     & Latitude     & Longitude & Time \\
    \midrule
    250173164003 & 41.83  & -72.237 & 1:00pm \\
    250173164003 & 41.84  & -72.243 & 1:01pm \\
    250173164003 & 41.83  & -72.251 & 1:01pm \\
    250173164003 & 41.82  & -72.257 & 1:03pm \\ \bottomrule
  \end{tabular}
  \label{tab:agg3}
\end{table}

This data can represent a precise distribution of people over space and time.  It can be used to observe the places where groups of people congregate, spend time, or how groups of people move through space.

\subsubsection{Market Utility}
A common use case of location data for physical retailers is to help plan new business locations or expand their customer base \cite{white_analytics}. Retailers can use aggregate location data to observe how much time people spend near a potential new storefront and compare foot traffic at nearby businesses and amenities.  

Businesses can also use this type of location data to estimate how far consumers are willing to travel for certain amenities or stores.  The identifying census block group that corresponds to each point can be used as a proxy for where someone lives, and how far someone traveled to a given point such as a store’s location can be estimated.

A potential consumer’s distance to a store has been an important variable in retail location theory, such as modeling the best placement for a new supermarket outlet \cite{baviera2016geomarketing}.
	
This data can also be used in more targeted marketing strategies.  For example, Athleta used location data to identify local areas of consumers who were more likely to become Athleta customers.  Athleta successfully expanded their customer base by using “neighborhood targeting” to drive visits from new customers in targeted neighborhoods \cite{groundtruth_athleta}.

\subsubsection{Public Utility}
Cities can use aggregate location data to better understand the behaviors and activities of their citizens \cite{ratti2006mobile} and plan urban interventions based on their real-time location dynamics. 

Demographers can also use this data to measure how different spaces are used by different socioeconomic groups.  For example, segregation is often measured based on where people live, but researchers at MIT have used aggregate location data to also represent segregation based on where people spend their time.  The Atlas of Inequality project (https://inequality.media.mit.edu)\cite{atlasofinequality_paper} shows the distribution of different socioeconomic groups at amenities throughout the city of Boston.  This project leverages aggregate location data by inferring the socioeconomic groups from the census block group connected to the precise points of the amenities people visit.  Analysis from this project is of use in city zoning and planning, as cities make efforts to develop more diverse environments.

\subsubsection{Risk}
Data aggregated at this level provides clear improvements to individual privacy, as trajectory data is no longer directly linked to an individual. However, with advanced methods, the likely trajectories of neighborhood residents can be estimated. Recent research has shown that in ideal scenarios, user trajectories can be recovered with up to 91\% accuracy from aggregated location data that was collected from mobile applications \cite{xu2017trajectory}.  

However, it’s also worth noting that re-identifying an individual’s trajectory from this aggregated data is complex and now takes an additional step: an attacker first needs to reconstruct trajectory data for individual users.  Then, after reconstructing trajectories, an attacker can match location points to a user record. The difficulty of this process likely scales with the number of users represented in each neighborhood, and the reconstructed trajectories are only probabilistic.  They lack the certainty of knowing one user’s point is followed by the next. These probabilistic trajectories can still provide useful information to aid predatory marketers and skip-tracing firms, and for that reason they present risk to the people who may be pursued.

\subsection{Coarsened Aggregated Location Data}
\subsubsection{Level of Aggregation}
This case of aggregate data is similar to the previous in that it does not represent individual user trajectories.  Location points generated by a user are linked to a census block group for the user, instead of a unique user ID.  The location points are also aggregated to a similar level of neighborhood or census block instead of specifying longitude and latitude, making this data lower resolution than the previous case.  (See Table \ref{tab:agg4}). This data again represents the distribution and movements of people over space and time but obscures the precise locations that were visited.

\begin{table}
 \caption{Example of coarsened aggregated location data. Each row details the home census block group of a user, the census block group they were detected as being in, and the time at the resolution of minutes.}
  \centering
  \begin{tabular}{llll}
    Home Census Block Group     & Visiting Census Block Group & Time \\
    \midrule
    250173164003 & 330150540002 & 1:00pm \\
    250173164003 & 330150540002 & 1:01pm \\
    250173164003 & 330150540002 & 1:15pm \\
    250173164003 & 330150540002 & 1:32pm \\ \bottomrule
  \end{tabular}
  \label{tab:agg4}
\end{table}

\subsubsection{Market utility}
Retailers can still use data at this level of spatial resolution to inform decisions for where to open new outlets.  They can observe the areas where people congregate at different parts of the day and how far those locations are from people’s home areas in order to predict the success of potential store locations. Businesses can also continue to use this aggregated location data to identify audiences, although their estimates will be less granular.   For example, sporting goods retailers might observe the areas from which people tend to travel to camp grounds and other outdoor activity locations, and focus marketing efforts to those areas. Marketing firms might observe the areas from which people travel to specific types of large events, like music festivals, in order to target marketing campaigns for similar events in the future.

\subsubsection{Public Utility}
Coarsely aggregated location data can be used to supplement census data and provide population density data to countries where conducting an accurate census is not feasible \cite{deville2014dynamic}.

Census data is an important tool for government agencies.  For example, in the United States, it is used for both governance and infrastructure.  U.S. Census data is used to allocate the seats of the U.S. House of Representatives to the states based on their population, and it is used to inform decisions on where to build and maintain schools, hospitals, transportation infrastructure, and police and fire departments.  However due to the cost and logistic complexity of conducting large surveys, the U.S. Census is only conducted every ten years, and can therefore only provide a snapshot of the population at a point in time.  In other countries, census data is not collected at all.  Recent studies demonstrate how location data collected by mobile phone network operators can provide cost-effective and accurate maps of population distributions over national scales and any time period \cite{deville2014dynamic}.  This data can fill in the gaps between the infrequent collections of census data, and provide population estimates in countries where census data is not available.

\subsubsection{Risk}
Data aggregated at this level presents significantly less risk than the previous levels.  Individuals are not directly linked to locations, and locations are recorded at a low spatial resolution. 

However, individual trajectories can still be inferred from this data.  The same researchers who reconstructed individual trajectories from aggregate location data generated by mobile applications did the same with aggregate location data provided by a wireless network operator \cite{xu2017trajectory}.  The location data provided by the network operator was of low spatial resolution, and then aggregated to remove unique user IDs, making it match the aggregation level discussed here. The researchers reported a 73\% accuracy rate for reconstructing individual trajectories from this data. 

Yet, even if advanced methods are used to reconstruct user trajectories from these types of aggregate datasets, the level of risk associated with them is still much lower. The reconstructed trajectories are only probabilistic and the points within them are of low spatial resolution. These inferred trajectories only represent the areas where a user \textit{probably was} at different points in time.
It's quite difficult to abuse this data to identify an individual using this method. Consider the case of an analyst trying to uncover a user's "full" trajectory from an aggregated dataset, using some external location data like geotagged tweets.  Of all the trajectories in the dataset, 27\% of them are likely to have been reconstructed incorrectly. At best, the analyst will receive a handful of \textit{likely} trajectories, 27\% of which are wrong, that are \textit{similar} to the one they are searching for. This data can still be abused, but with far less precision than the more granular datasets we discuss.

\begin{figure}
  \centering
  \includegraphics[scale=.75]{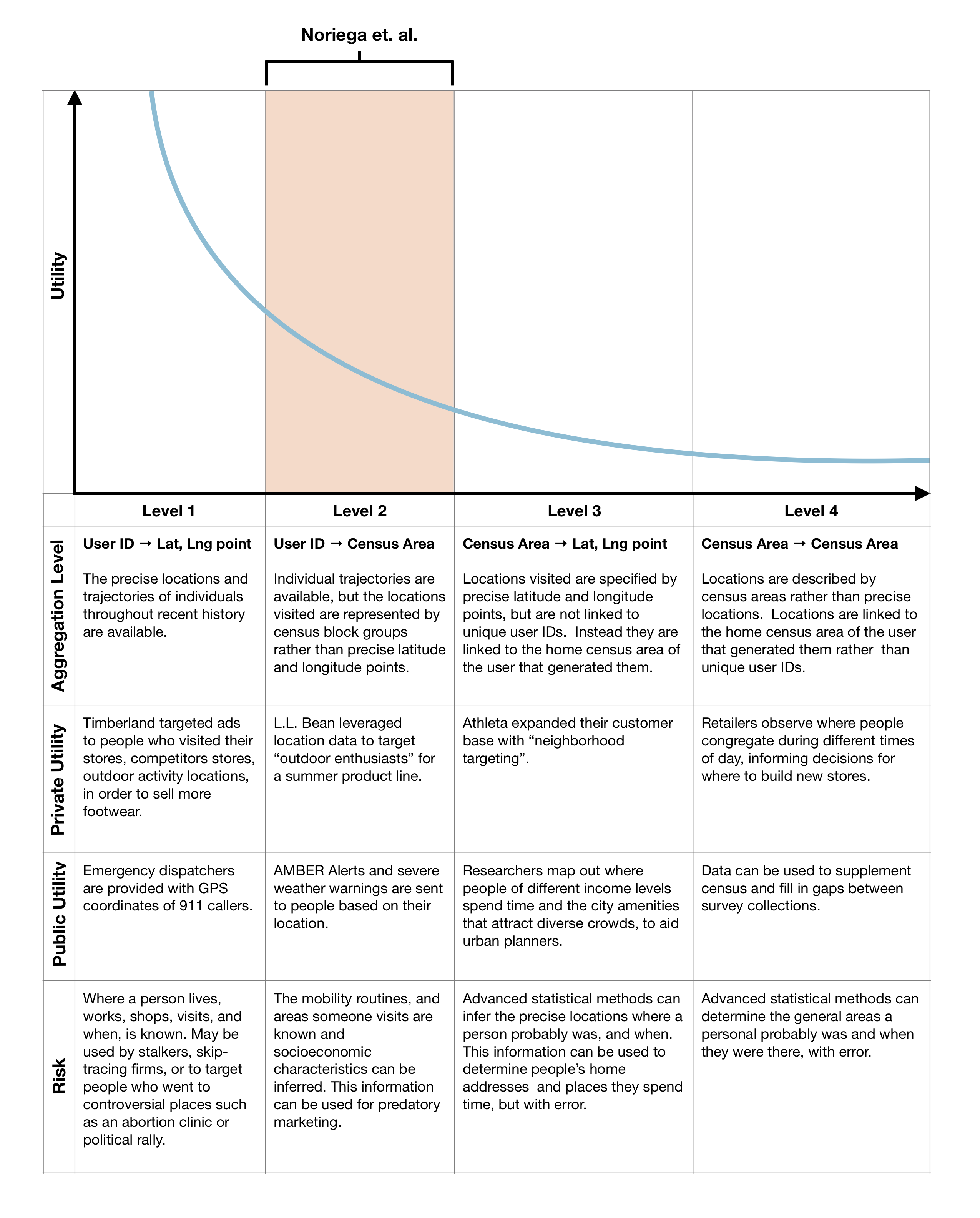}
  \caption{A graphical summary of location data aggregation strategies, with examples for the use of data aggregated at each level, and the data’s associated risk.  At each level  of aggregation, we provide examples of how the data can serve both private and public utility.}
  \label{fig:curve-table}
\end{figure}

\section{Discussion}
Data detailing human mobility at scale has a wide array of uses and utility, but carries a significant personal risk. Researchers who attempt to measure and describe this relationship between risk and utility often define both as quantitative attributes of data, rather than through the contextual impacts of its use. Quantitative measures of risk and utility do provide a useful compass and starting point: they can help researchers and policymakers understand how aggregation and anonymization strategies affect the value of location data as a traded commodity. However, when the real-world usage and risks of various data aggregation levels are considered, it becomes clear that risk and utility are closely coupled with the interests of who is using the data (e.g. public good vs private firms) and the people detailed in these datasets who are most at risk.

By examining example use-cases of location data by both private and public entities, some common patterns emerge. Many private uses of high-resolution location data provide value through identifying information at some resolution about individuals. Even if the data describes indirect information about an individual, being able to link it to a particular advertising target or profile is crucial in current value models. For instance, knowing the behavioral profile of people that live in a certain neighborhood or census block group provides firms with a greater contextual understanding of \textit{individuals who live in those areas}, not the areas themselves. In the cases of skip-tracing firms and individualized marketing, the identifiability of the data is it’s selling point, and so the value of the products hinge on their resolution at the individual level.

In contrast, many public use cases consist of identifying aggregate trends, such as measuring mobility patterns between neighborhoods or augmenting census data with inferred demographics. Cases in which identifiable information is needed to provide a public service, such as the AMBER alert system or e911, do not generally depend on storing location patterns. An exception to this general rule is location data use by law enforcement, the carceral system, and ICE, where location data’s largest value stems from targeting and tracking individuals \cite{aclu2019}.

Current research on the privacy and risk of location data often addresses strategies for mitigating re-identifiability risk while maintaining its ‘utility’. This discourse generally assumes that data providers, those collecting and sharing the data, have an incentive to balance re-identifiability risk with its usefulness, but this is not the case. In most use cases in the private market, the ability to link LBS data to other external datasets, like online behavioral profiles or real-world identities, is the source of its primary value. The current discourse also identifies risk and utility as \textit{static} attributes of data.  In reality, both are dynamic quantities that vary as technology advances and additional datasets become available that can be used to reconstruct trajectories\cite{brickell2008cost,li2009tradeoff,noriega2018mapping}. A prime example of this can be found in recent work by Fengli Xu et. al., where researchers adversarially reconstruct likely individual trajectories from highly aggregated location data \cite{xu2017trajectory}. 
This example is particularly concerning because their aggregated dataset would have previously been considered privacy-preserving. While the impact of this development on the risk of location data seems to be small, this may not be true in the future. It is impossible to know what novel techniques for re-identification may be developed, and so the risk in releasing a location dataset can be estimated, but not known.

There are some projects that propose strategies other than data aggregation that allow for the use of location and other sensitive datasets while providing firm privacy guarantees \cite{barak2016anonymizing,tu2017beyond,hardjono2019trusted,gramaglia2014anonymizability}. These projects include proposals to change how parties currently manage and access user data, such as the Safe Answers framework and its derivatives, where value is extracted from data by running vetted algorithms on black-box data stores that individuals control \cite{de2014openpds,hardjono2019trusted}. 
While we believe that new data-sharing and ownership models are crucial to the future of data as a good generally, many will likely require a large normative change in how data markets function. For example, the Safe Answers framework ostensibly requires that data collection happens in a decentralized manner, a requirement that is at odds with our current laws of information goods and the highly centralized way that applications are built today \cite{spiekermann2015challenges}. 

As it stands today, there is an imbalance between how LBS data is made available to and used by private firms versus public interest groups. LBS data is collected from the public, yet primarily generates value for private firms, and there are no regulatory requirements or agreements in place that require it be shared. However, there is precedent for data generated by or with the help of the public to be shared freely with public interest groups. Many private companies, such as city bike share companies, subway contractors, and telecom companies, sign data sharing agreements in order to operate \footnote{For example, New York City has an open data portal (https://opendata.cityofnewyork.us) and mission statement “...We believe that every New Yorker can benefit from Open Data, and Open Data can benefit from every New Yorker.”}. What makes these examples different from the location data collected by LBS firms?

One possible key difference is the perceived use of infrastructure. U.S. telecom and internet companies use cell towers built through public-private partnerships that the federal government subsidizes heavily \cite{crawford_2018}. Bike sharing programs like New York’s CitiBike or Boston’s Bluebikes are often owned by the cities themselves as public infrastructure, but operated by a private firm that is required to share usage data with the city \cite{levy_2018}. Ride-sharing companies like Uber and Lyft use public roadways to operate, and have recently faced legal pressure to enter into data sharing agreements with cities. Lyft, for example, was recently ordered to share data with the city of Seattle by way of its public records law, as the data was found to be of clear value to public interest  \cite{lyft.v.seattle_2018}. Michael Ryan, a Seattle assistant city attorney, explained that Lyft is “a business that is conducted solely on the city streets and the taxpayers pay for the streets” \cite{gutman_2019}. These examples demonstrate a precedent for private firms that use public infrastructure to share the data that is collected via that infrastructure.  Yet in all these examples, a transaction between public infrastructure and public data collected by private firms is clear.

The “infrastructure” used by LBS companies is far more muddled than in these cases. LBS companies operate by providing convenient ways for firms to access a user’s location from their smartphone, but do not utilize GPS services or telecom services directly like telecom companies. They also do not facilitate urban movement directly, like in the case of bike sharing or ride sharing firms. Even so, the data is still generated \textit{by the public}, and like the example of the Seattle case against Lyft, there is a strong case to be made that aggregated LBS data is in the public interest. Without a clear transaction of public infrastructure, can private firms be compelled to share data collected about the public?

If this data is to be shared, then deciding how it is shared is also important. Ride sharing services often share data in the form of aggregated trips between census tracts. This is similar to our second aggregation case, where user trajectories are coarsened to the level of census statistical areas. However, the risk profile of ridesharing data is very different than location service data because it is less ubiquitous. First, only individuals that use or can afford ridesharing services are represented in ridesharing data. Second, only fragments of individual mobility patterns are represented in data from ridesharing services. This is in contrast to the more pervasive data collected by LBS firms, whose resulting datasets are more complete, and can include a user’s location before, during and after a ride. Still, like ridesharing services, aggregation is the clearest way forward for LBS firms to share data with the public given the current data market. These considerations only begin to address the \textit{imbalance} in the current public and private markets for location data. They do not address the significant \textit{risk} involved in the current unregulated data market, where individual trajectories are sold and purchased with little oversight.

How to manage this risk, let alone open the value of LBS data to the public remain open questions.  Until these questions are answered, an imbalance between individual risk and the private and public utility presented by LBS data will persist.

\bibliographystyle{IEEEtran}  
\bibliography{references}

\end{document}